# Enchanced reflectance SiNx/SiOx DBR mirror based on TEOS precursor fabricated by PECVD method.

Asharchuk I.M.[1,2], Shibalov M.V.[1], Mumlyakov A.M.[1], Nekludova P.A.[1], Diudbin G.D.[1], Zenova E.V.[1], Minaev N.V.[2], Pavlov A.A.[1] and Tarkhov M.A.[1]

[1] Institute of Nanotechnology of Microelectronics of the Russian Academy of Sciences, Moscow 119991, Russia

[2] Institute of Photon Technologies, Federal Scientific Research Centre 'Crystallography and Photonics', Russian Academy of Sciences, Troitsk, Moscow 108840, Russia

E-mail: ilyaasharchuk@gmail.com

**Abstract**

In this study, we investigated the influence of silicon oxide roughness produced from the gas precursor monosilane (SiH4) and the silicon-organic precursor tetraethoxysilane (TEOS) on the optical qualities of a distributed Bragg reflector (DBR). A significant influence of the precursors from which SiOx is deposited on the optical qualities of DBR mirrors is demonstrated in this study. It has been shown experimentally that a mirror produced from the TEOS precursor is endowed with better qualities than a mirror produced using SiH4, that is, the reflectivity increased by 20% and optical losses decreased by half. The roughness average (Ra) of the SiNx/SiOx / (TEOS) mirror surface decreased by a factor of five compared to that of the SiNx/SiOx / (SiH4) mirror.

**Introduction**

The quality of the dielectric Bragg mirrors is of prime importance for various applications in modern photonics and applied laser physics. Generally, two factors affect the quality: scattering loss caused by the roughness in thin-layer interfaces and material absorption loss [1]-[5]. For instance, losses related to light scattering may affect the threshold of laser damage in thin films. Backscattering is critical for layered films with high reflectivity implemented in measuring systems where lasers are used [6]–[7]. As long as the layer thickness of such mirrors is dozens of nanometers, and they are transparent in the visible light spectrum of 400-750 nm, the absorption coefficient is rather low (400 cm$^{-1}$) and the approach to loss reduction is to improve the quality of the surface between the interfaces of the film pairs.

Currently, the development of single quantum emitters based on synthesized diamonds with NV centers is a relevant issue [8]. Optical pumping of the emitters was performed at a wavelength of 532 nm. Therefore, the development of DBR mirrors for this wavelength band is important. It is reliably known that the quantum efficiency (QE) of single-photon detectors is increased by integrating a sensing element into an optical cavity based on DBR [9]. DBR mirrors are based on the layer-by-layer deposition of dielectric films with specific refractive indices. The most affordable dielectric pair was SiN$_x$/SiO$_x$. According to F. Reveret et al. [1], when HfO$_2$ is used alternatively to SiN$_x$, the coating roughness Ra is decreased by a factor of 2.5 and is equal to ~ 1 nm. However, the use of HfO$_2$ is impractical because of the high cost of materials. DBR mirrors based on the $SiN_x/SiO_x$ dielectric pair function properly over a wide range of wavelengths, and the silicon-containing precursors used for layer deposition are affordable. As an affordable precursor for producing $SiO_x$ and $SiN_x$ coatings, monosilane ($SiH_4$) is implemented. The deposition of several layers of the $SiN_x/SiO_x$ pairs was carried out in one technological chamber in situ.

The data on the roughness of films used as dielectric pairs of a mirror optimized for UV radiation were given by Dai et al. [10]. The roughness of these films was $R_a$ ~ 2.93 nm - 4.34 nm. We assume that a decrease in the film roughness is possible when using tetraethylorthosilicate (TEOS) as a precursor to form the $SiO_x$ layer. It has been established that $SiO_x$ deposited from the TEOS precursor can achieve a roughness $R_a$ of ~ 1.5 nm [11]. The roughness values of some layers of the DBR mirror significantly affect the value of the scattering at the interface of the $SiN_x/SiO_x$ dielectric pairs [12], [13].

In this study, the effect of silicon oxide produced from various precursors ($SiH_4$ and TEOS) using the PECVD technique on the optical properties of multilayer dielectric mirrors based on the sequence of the $SiN_x/SiO_x$ pairs was experimentally investigated.

**Experimental setup and methods**

By alternating the $SiN_x/SiO_x(SiH_4)$ and $SiN_x/SiO_x(TEOS)$ layers in pairs, dielectric mirrors were produced on sapphire and amorphous silicon substrates and optimized for a wavelength of 540 nm. The $SiN_x/SiO_x$ dielectric pairs were deposited in seven quantities in a technological chamber without vacuum discontinuities. A layer of $SiN_x$ was deposited from a mixture of monosilane (100% $SiH_4$), ammonia ($NH_3$), and nitrogen ($N_2$) using the PECVD technique. The pressure in the technological chamber is 166.65 Pa. The power of the RF source was set at 30 W. A silicon oxide layer was deposited from two different silicon-containing precursors using the PECVD technique. In the former case, $SiO_x$ was deposited from a gas mixture of $SiH_4$ and nitrogen oxide ($N_2O$). In the latter case, silicon oxide was deposited using a gas mixture of tetraethoxysilane (TEOS) and oxygen ($O_2$). The pressure in the chamber is 53.33 Pa. The power of the RF source was 70 W and the stage temperature was 300°C for all the deposition processes.

Atomic force microscopy (AFM) measurements were carried out to assess the root-mean-square roughness ($R_a$) and waviness ($W_a$) of the surfaces of various layers. An area of 50×50 μm was selected to analyze the waviness $W_a$ of the mirror surface, whereas areas of 1×1 μm were used to measure the roughness.

The spectra of transmission and reflectivity at a normal angle of incidence were measured using an Ocean Optics HL-2000 halogen light source coupled with a Thorlabs fiber. The fiber diameter was 400μm, the fiber numerical aperture (NA) was 0.39. The radiation was collimated from the fiber using an Ocean Optics 74-UV collimating lens. The spectra were detected using a Fluorolog-3 spectrofluorometer (Horiba Jobin Yvon, France) equipped with an iHR 320 monochromator and a Hamamatsu photomultiplier tube R928. The refractive indices of the thin films were measured by ellipsometry. The indices of refraction were 1.950, 1.470 and 1.458, for $SiN_x$, $SiO_x(SiH4)$, and $SiO_x(TEOS)$, respectively. All the measurements were performed at room temperature.

**Results and discussion**

As it can be seen from AFM images given in Figure 1, the average value of roughness $R_a$ of the $SiN_x/SiO_x$ ($SiH_4$) mirror is 11.04 nm, and $SiN_x/SiO_x$ (TEOS) – 2.31 nm. The values of average waviness $W_a$ for the two mirrors are almost ten times different, and constitute 202 pm for the mirror produced with the $SiH_4$ precursor and 35 pm for the TEOS-produced mirror. It is worth noting that the average waviness and roughness values depended on the precursor used. Surface roughness and waviness are directly associated with the roughness of the interface in the DBR cutoff.

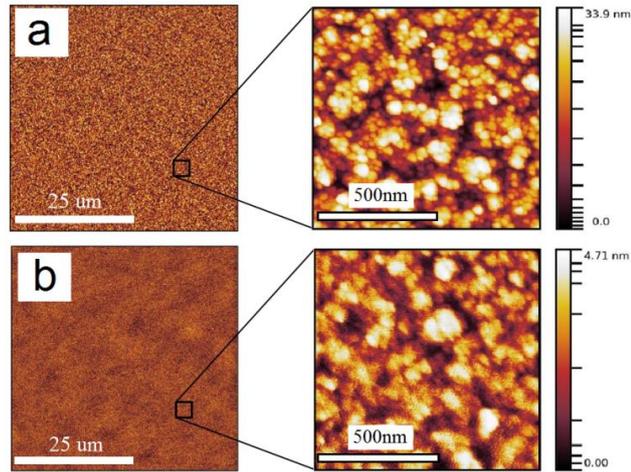

Figure. 1. AFM images of the top surface of a 7 pair (a) $SiN_x/SiO_x$(TEOS) and (b) $SiN_x/SiO_x$ ($SiH_4$)

A scanning electron microscopy (SEM) image of the DBR mirror cutoff, that is, $SiN_x/SiO_x$ (TEOS), including the representation of a sequence of layers of dielectric pairs, is shown in Figure 2(a) (lighter layers $SiO_x$, darker layers $SiN_x$, Figure 2 (b, c) ). The element distribution map obtained using energy dispersive X-ray analysis (EDAX) is shown in Figure 2 (b, c). The image shows the distribution of nitrogen and oxygen at the mirror cutoff, corresponding to silicon and nitride oxides, respectively.

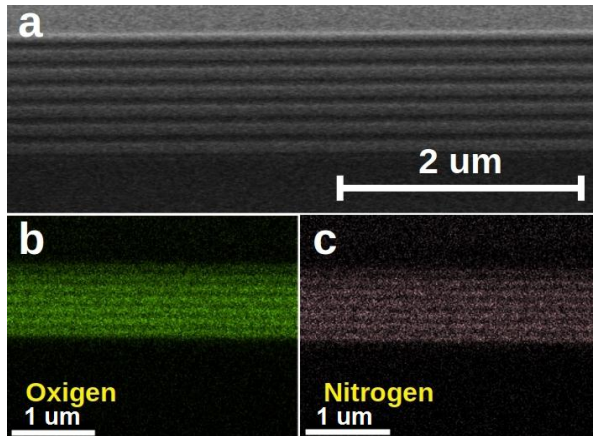

Figure. 2. (a) SEM image of the DBR mirror based on 7 pair $SiN_x/SiO_x$(TEOS). (b) Color map of oxygen distribution (c) Color map of nitrogen distribution

The reflectance band and reflectivity are the basic optical properties that distinguish the produced mirrors. It is possible to determine the reflectance band based on the matrix method for radiative transfer as follows:

$$\Delta\lambda = \frac{\lambda_c}{\pi} arcsin\left(\frac{n_{high}-n_{low}}{n_{high}+n_{low}}\right) \quad (1)$$

where $\lambda_c$ is a mid-length of the band $\Delta\lambda$ determined by the period d of a double layer by Bragg law:

$$\lambda_c = 2dn_{eff}, n_{low} < n_{eff} < n_{high} \quad (2)$$

where $n_{eff}$ is the effective refractive index and $n_{high}$ and $n_{low}$ are the refractive indices of the layers with higher and lower refractive indices, respectively, composing the bilayer. The reflectivity is determined as [14]

$$R = \left[\left(\frac{n_{high}}{n_{low}}\right)^{2N} - 1 \bigg| \left(\frac{n_{high}}{n_{low}}\right)^{2N} + 1\right]^2 \quad (3)$$

where N is the number of periods (double-layer). The values specified above depend on the refractive index difference of the $SiN_x/SiO_x$ samples.

The transmission and reflectivity spectra of the dielectric mirrors produced on sapphire substrates using the $SiH_4$ and TEOS precursors are shown in Figure 3 (a) and (b), respectively.

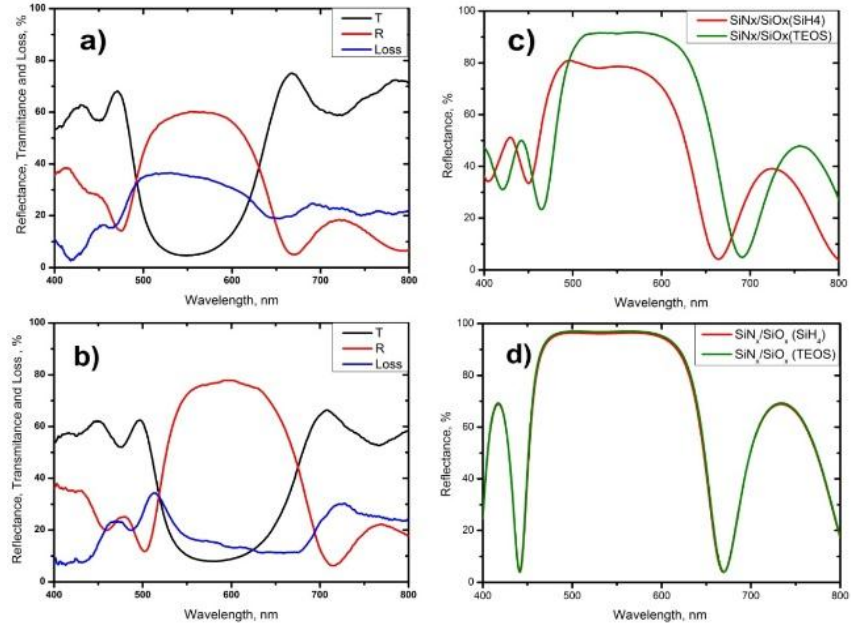

Figure 3. a) reflectivity, transmission, and calculated losses of 7 pairs $SiN_x/SiO_x$ ($SiH_4$) DBR on the sapphire substrate; b) reflectivity, transmission, and calculated losses of 7 pairs SiNx/SiOx(TEOS) DBR on the sapphire substrate; c) reflectivity of 7 pairs $SiN_x/SiO_x$ (TEOS), green curve, and 7 pairs SiNx/SiOx($SiH_4$) DBR, red curve - on the silicon substrate; d) Reflection spectra calculated for 7 pairs $SiN_x/SiO_x$ (TEOS) and $SiN_x/SiO_x$ ($SiH_4$) on the silicon substrate.

The reflectivity of $SiN_x/SiO_x$ (TEOS) is 20% higher than that of $SiN_x/SiO_x$ ($SiH_4$) (Figure 3 (a, b)). Considering Equation (3) for the reflectivity calculation, it can be observed that the refractive index difference of the $SiN_x/SiO_x$ pairs for mirrors produced using different precursors but with the same technology cannot facilitate the same increase in reflectivity as that shown in Figure 3 (a, b, c). We assume that the morphology of the surface between the deposited layers of the distributed Bragg mirror significantly influences its optical properties. In this study, we used a simple model to determine the transmission coefficients of seven pairs of $SiN_x/SiO_x$ (TEOS) and $SiN_x/SiO_x$ ($SiH_4$) DBR layers (Figure 3 d). In this model, we neglected the substrate's absorption and assumed that the interfaces between the layers tended to a perfectly even surface. An insignificant difference in reflectivity for $SiO_x(SiH_4)$ and $SiO_x(TEOS)$, equal to 1.470 and 1.458, respectively, is evident. Because the values of the absorption coefficients for $SiN_x$ and $SiO_x$ are rather low in the visible-light spectrum [15], the effect of absorption in the films can be neglected. Only one remaining component affects the reflectivity and transmission coefficients, that is, scattering at the interfaces.

The experimental results are shown in Figures 1, and 3 support our assumption. The loss reduction is shown in Figure 3 (b). The loss value decreased twice for DBR produced using TEOS. The loss values were calculated using the following formula: *Loss = 1- R-T* [16]. The reflectivity of the mirrors produced on silicon substrates was 15% higher than that of the mirrors produced on sapphire substrates. This is because the reflectivity of polished silicon is ten times higher than that of sapphire $R_{Si}$, that is, ~ 30-35% at a wavelength of 540–532 nm [17].

For the sample produced using the TEOS precursor, a shift to the long-wave region was observed. This shift can be attributed to the thickening of layers during production. According to the Bragg law formula used to determine the resonant wavelength, the resonance position depends on the effective refractive index and period. As far as the refractive index difference $n$ of the $SiO_x(SiH_4)$ and $SiO_x$ (TEOS) layers is equal 0.012 (1.47-1,458=0.012) there is no shift of the central wavelength according to the theoretical curve of the reflectivity (Figure 3 d), and the influence of $n_{eff}$ value on the shift $\lambda_c$ can be neglected. Therefore, the change in the period considerably contributes to the shift in the resonant wavelength of the Bragg mirror.

**Conclusion**

Comparing the two types of mirrors produced using different precursors demonstrates that the implementation of the TEOS precursor increases the quality of the interfaces of the deposited layers, which results in the growth of the reflectivity of DBR by 20% with respect to DBR produced with SiO2 from silane precursors, and reaches a maximum value close to 80% at 600 nm. Further, the surface roughness average $R_a$ from 11.04 to 2.31 nm and substantially decrease the waviness $W_a$ value substantially decreases from 202 to 35 pm. As a result, the $SiN_x/SiO_x$ interface improvement leads to a loss reduction of two times. These results can be interpreted as a significant advantage, contributing to a reduction in the material costs for the production of multilayer mirrors by reducing the processing time owing to the small number of pairs of layers and the improvement in the quality required to achieve the target optical properties of the dielectric mirror.

**Acknowledgements**

This work was supported by GK №20411.1950192501.11.003 Ministry of Industry and Trade in part of design and develop of DBR mirror; Ministry of Science and Higher Education within the State assignment FSRC 'Crystallography and Photonics' RAS in part of measurement optical properties of DBR.

**References**

[1]     F. Réveret et al., "High reflectance dielectric distributed Bragg reflectors for near ultra-violet planar microcavities: SiO 2 /HfO 2 versus SiO 2 /SiN x," J. Appl. Phys., vol. 120, no. 9, p. 093107, Sep. 2016, doi: 10.1063/1.4961658.
[2]     M.-J. Choi, O. D. Kwon, S. D. Choi, J.-Y. Baek, K.-J. An, and K.-B. Chung, "Enhanced Anti-reflective Effect of SiN x /SiO x /InSnO Multi-layers using Plasma Enhanced Chemical Vapor Deposition System with Hybrid Plasma Source," Appl. Sci. Converg. Technol., vol. 25, no. 4, pp. 73–76, Jul. 2016, doi: 10.5757/ASCT.2016.25.4.73.
[3]     I. Ohlídal, "Approximate formulas for the reflectance, transmittance and scattering losses of nonabsorbing multilayer systems with randomly rough boundaries," J. Opt. Soc. Am. A, vol. 10, no. 1, p. 158, Jan. 1993, doi: 10.1364/JOSAA.10.000158.
[4]     K. H. Guenther, H. L. Gruber, and H. K. Pulker, "Morphology and light scattering of dielectric multilayer systems," Thin Solid Films, vol. 34, no. 2, pp. 363–367, May 1976, doi: 10.1016/0040-6090(76)90492-2.
[5]     H. K. Pulker, "Optical losses in dielectric films," Thin Solid Films, vol. 34, no. 2, pp. 343–347, May 1976, doi: 10.1016/0040-6090(76)90483-1.
[6]     Y. Pan, Z. Wu, L. Hang, and Y. Yin, "Light scattering losses of high reflection dielectric multilayer optical devices," Thin Solid Films, vol. 518, no. 8, pp. 2001–2005, Feb. 2010, doi: 10.1016/j.tsf.2009.08.007.

[7]     H. Hou, K. Yi, S. Shang, J. Shao, and Z. Fan, "Measurements of light scattering from glass substrates by total integrated scattering," Appl. Opt., vol. 44, no. 29, p. 6163, Oct. 2005, doi: 10.1364/AO.44.006163.

[8]     Trupke M., Munro W., Nemoto K., et. al. "Enhancing photon collection from quantum emitters in diamond", Progress in Informatics, Vol. 8. 2011, doi: 10.2201/NiiPi.2011.8.4

[9]     Zhang, W., et al. (2017). "NbN superconducting nanowire single photon detector with efficiency over 90% at 1550 nm wavelength operational at compact cryocooler temperature ". Science China Physics, Mechanics & Astronomy,60(12), 1-10. doi: 10.1007/s11433-017-9113-4

[10]    J. Dai et al., "Design and fabrication of UV band-pass filters based on SiO2/Si3N4 dielectric distributed bragg reflectors," Appl. Surf. Sci., vol. 364, pp. 886–891, Feb. 2016, doi: 10.1016/j.apsusc.2015.12.222.

[11]    C. E. Viana, A. N. R. da Silva, N. I. Morimoto, and O. Bonnaud, "Analysis of SiO2 Thin Films Deposited by PECVD Using an Oxygen-TEOS-Argon Mixture," Brazilian J. Phys., vol. 31, no. 2, pp. 299–303, Jun. 2001, doi: 10.1590/S0103-97332001000200023.

[12]    I. Ohlidal, "Approximate formulas for the reflectance, transmittance, and scattering losses of nonabsorbing multilayer systems with randomly rough boundaries", J. Opt. Soc. Am. A, 1993, 10(1), doi: 10.1364/jossa.10.000158

[13]    I. Ohlidal. "Reflectance of multilayer systems with randomly rough boundaries", Optics Communications, Vol. 71, no. 6, pp. 323-326, Jun. 1989, doi: org/10.1016/0030-4018(89)90041-2

[14]    H. A. Macleod, Thin-Film Optical Filters. CRC Press, 2010